\documentclass[twocolumn,prl,showpacs,superscriptaddress]{revtex4}

\usepackage{amsmath,amssymb,amsfonts}
\usepackage{CJK}
\usepackage{bm}
\usepackage{array}
\usepackage{multirow}

\begin{document}

\title{Topological Classification and Stability of Fermi Surfaces}


\author{Y. X. Zhao}
\email[]{zhaoyx@hku.hk}
\author{Z. D. Wang}
\email[]{zwang@hku.hk}

\affiliation{Department of Physics and Center of Theoretical and Computational Physics, The University of Hong Kong, Pokfulam Road, Hong Kong, China}


\date{\today}

\begin{abstract}
In the framework of the Cartan classification of Hamiltonians, a kind of topological classification of
Fermi surfaces is established in terms of topological charges. The topological charge of a Fermi surface
depends on its codimension and the class to which its Hamiltonian belongs. It is revealed that six types of
topological charges exist, and they form two groups with respect to the chiral symmetry, with each group
consisting of one original charge and two descendants. It is these nontrivial topological charges which
lead to the robust topological protection of the corresponding Fermi surfaces against perturbations that
preserve discrete symmetries.\end{abstract}

\maketitle


For a Fermi gas at zero temperature,  a Fermi surface(FS) naturally arises as the boundary separating
the occupied  and empty states in $(\omega,\mathbf{k})$-space.
As is known, when  weak interactions/perturbations
and disorders are introduced, some FSs still survive though the occupation
of states may be shifted dramatically, while some others are easily
gapped. Such kind of FS stability originates from its
topological property of the Green's function or the Feynman propagator
for fermionic particles, $G(\omega,\mathbf{k})=\left(i\omega-\mathcal{H}\right)^{-1}$,
which was first pointed out by Volovik in Ref.\cite{Volovik-Book}.
Generally speaking, an FS is robust against weak interactions/perturbations and disorders, if it has a nontrivial topological charge that
provides the protection; otherwise it is vulnerable and easy to be gapped.
The most general case is that a Hamiltonian is not subject to any
symmetry, where the topological charge is formulated by the homotopy
group $\pi_{p}(GL(N,\mathbf{C}))$. This general case was addressed
in Refs.\cite{Volovik-Book,Volovik Vacuum} and analyzed
 in the framework of the K theory \cite{Horava}. Notably, real physical systems
have normally certain symmetries, making it necessary and significant to develop
a corresponding theory for symmetry preserving cases.

It is known that the symmetry
of a quantum system can always have either a unitary representation
or an anti-unitary one in the corresponding Hilbert space. The unitary symmetries, such as rotation, translation, and parity symmetries, are easy to be broken by weak interactions/perturbations and disorders, while the anti-unitary symmetries, such as the time-reversal symmetry(TRS) and charge
conjugate or particle-hole symmetry(PHS)  can usually be preserved. Thus we are motivated to
develop a unified theory to classify the FSs of systems with the two so-called reality symmetries, namely TRS and PHS, which seems to be fundamentally important.
In this Letter, taking into account the two reality symmetries and introducing six types of topological charges,  a new kind of complete classification of all FSs is obtained, being explicitly illustrated in Tab.{[}\ref{tab:periodicity table}{]}. Moreover, an intrinsic relationship between the symmetry class index and the codimension number is established, which provides us a unique way for realizing any type of FS with a high codimension.




Let us first introduce how to classify all sets of
Hamiltonian involving the above-mentioned two reality symmetries, as done in the random matrix theory\cite{Ramdom Matrix 0,Ramdom Matrix I}. It turns out that the chiral symmetry(CS) has to be introduced for a
complete set of classification.
If a unitary symmetry represented by $\mathcal{K}$ can anti-commute
with the Hamiltonian $\mathcal{H}$, $i.e.$, $\left\{ \mathcal{K},\mathcal{H}\right\} =0$,
it corresponds the CS. In fact, the combined symmetry of TRS and PHS
is a kind of chiral symmetry. On the other hand, two chiral symmetries can be combined to commute
with $\mathcal{H}$, which makes $\mathcal{H}$ diagonal. Thus it
is sufficient to consider only one chiral symmetry. Based on these considerations\cite{Ramdom Matrix II,TI Classification III,TI Classification I},
a kind of complete classification of all sets of Hamiltonian can be obtained with
respect to the three symmetries. As is known, the TRS and PHS are both anti-unitary,
 and can  be expressed in a unified form
\begin{equation}
\mathcal{H}(\mathbf{k})=\epsilon_{c}C\mathcal{H}^{T}(-\mathbf{k})C^{-1},\quad CC^{\dagger}=1,\quad C^{T}=\eta_{c}C,\label{eq:reality transformation}
\end{equation}
where $\epsilon_{c}=\pm1$ denotes TRS/PHS, and $\eta_{c}=\pm1$. As a result, each reality
symmetry may take three possible types (even, odd, and absent), and thus there
are nine classes. In addition,  considering that the chiral symmetry may be preserved or not
when both reality symmetries are
 absent,  we can have
 ten classes, as summarized in Tab.{[}\ref{tab:classification}{]},
which is the famous Cartan classification of Hamiltonians in the random matrix theory\cite{Ramdom Matrix 0,Ramdom Matrix I}, while may be easily understood in the present framework \cite{TI Classification I,TI Classification III,Ramdom Matrix II,Tl Classification II}.

\begin{table}
\begin{tabular}{|>{\centering}p{0.5cm}|>{\centering}p{0.6cm}>{\centering}p{0.6cm}>{\centering}p{0.6cm}>{\centering}p{0.6cm}>{\centering}p{0.6cm}|>{\centering}p{0.7cm}>{\centering}p{0.7cm}>{\centering}p{0.7cm}>{\centering}p{0.6cm}>{\centering}p{0.6cm}|}
\hline
 & \multicolumn{5}{c|}{Non-chiral case} & \multicolumn{5}{c|}{Chiral case}\tabularnewline
\hline
\hline
 & A & AI & D & AII & C & AIII & BDI & DIII & CII & CI\tabularnewline
\hline
T & 0 & +1 & 0 & -1 & 0 & 0 & +1 & -1 & -1 & +1\tabularnewline
\hline
C & 0 & 0 & +1 & 0 & -1 & 0 & +1 & +1 & -1 & -1\tabularnewline
\hline
S & 0 & 0 & 0 & 0 & 0 & 1 & 1 & 1 & 1 & 1\tabularnewline
\hline
\end{tabular}
\caption{Classification of Hamiltonian. T and C denote respectively TRS and PHS, with $0$ indicating
the absence of TRS(PHS) and $\pm1$ denoting the sign of TRS(PHS). S denotes CS, with $0$ and $1$ representing the absence and presence of CS. \label{tab:classification}}
\end{table}

For a system with spatial dimension $d$, the FS is a compact submanifold
with dimension $d-p$ if the $(\omega,\mathbf{k})$-space is compact. We may define the codimension of the FS
as $p$.  Topological charges can be defined on a chosen $p$-dimensional submanifold in the $(\omega,\mathbf{k})$-space enclosing the FS in its transverse dimension. The two reality symmetries are essentially different from the CS, which lies in that each of the reality symmetries relates a $\mathbf{k}$ point
to the $-\mathbf{k}$ point in $\mathbf{k}$-space,
while the CS acts on every $\mathbf{k}$. This difference leads
to the requirement that a chosen submanifold in $(\omega,\mathbf{k})$-space
has to be centrosymmetric around the origin in order to preserve either of the
reality symmetries.

Generally, the topological property of an FS depends on its codimension and symmetry~\cite{note1}. After the detailed analysis, we are able to classify all classes in an
appropriate form illustrated in Tab.{[}\ref{tab:periodicity table}{]}, which is one of main results of this work. In each case, as seen in
Tab.{[}\ref{tab:periodicity table}{]}, a kind of topological charge is designated to the FS. Topological charges $0$, $\mathbf{Z}$,  $\mathbf{Z}^{(1,2)}_{2}$, and $2\mathbf{Z}$
 correspond respectively to $0$, an integer,  an integer of $\mod2$, and  an even integer. As illustrated in the table, the ten classes can be divided into real and complex cases according to whether or not they have either of the reality symmetries. On the other hand, they can also be put into the chiral and non-chiral cases depending on whether or not they have the CS. Actually, from this kind of classification, we can find that there are six types of topological charges, which form two groups in terms of the CS, with each group consisting of an original one ($\mathbf{Z}$) and the two descendants ($\mathbf{Z}^{(1)}_2$ and $\mathbf{Z}^{(2)}_2$) \cite{note}.  In the chiral or non-chiral real case, all of the four types of FSs can be realized for any given number of codimension.  It is also clear that the complex and real cases have different periodicities: one is of a two-fold periodicity, while the other is of an eight-fold one. Obviously, for the complex case, we can introduce a matrix element $C(p,i)$ to denote the topological charges listed in Tab.{[}\ref{tab:periodicity table}{]}, where $i=$ odd(even) number corresponds to the class A (AIII), and $p$ is the number of codimension. In this way,  $C(p,i)$ satisfies the relation $C(p,i)=C(p+n,i+n)$ with $n$ an integer and $i\mod 2$. For the real case, we can also introduce another matrix element $R(p,i)$, where $i=1, ...,8$ denote the classes AI, BDI, D, DIII, AII, CII, C and CI respectively. Intriguingly, we can also find
 \begin{equation}
 R(p,i)=R(p+n,i+n)
 \end{equation}
with $i\mod 8$, which establishes an intrinsic relationship between the symmetry class index and the codimension number. More significantly,  based on this relationship, we are able to realize any type of FS with a high codimension by reducing its codimension to an experimentally accessible one with the same CS, $i.e.$, $p=1, 2 , 3$. Actually, the above relation and the two-fold periodicity as well as the eight-fold one are originated from the Bott periodicity of $GL(N,\mathbf{C})$. In particular, the present eight-fold periodicity  is induced by the two reality symmetries enforcing on the two-fold Bott periodicity, and thus is different from the eight-fold Bott periodicity for $O(N)$ and $Sp(N)$\cite{Nakahara}.

\begin{table}
\begin{tabular}{|>{\centering}p{0.6cm}|>{\centering}p{0.6cm}>{\centering}p{0.6cm}>{\centering}p{0.6cm}>{\centering}p{0.6cm}|>{\centering}p{0.7cm}>{\centering}p{0.7cm}>{\centering}p{0.7cm}>{\centering}p{0.7cm}|}
\hline
 & \multicolumn{4}{c|}{Non-chiral case} & \multicolumn{4}{c|}{Chiral case}\tabularnewline
\hline
\hline
 & \multicolumn{8}{c|}{Complex case}\tabularnewline
\hline
 & \multicolumn{4}{c|}{A} & \multicolumn{4}{c|}{AIII}\tabularnewline
\hline
$p\diagdown i$ & \multicolumn{4}{c|}{$\mathbf{1}$} & \multicolumn{4}{c|}{$\mathbf{2}$}\tabularnewline
\hline
1 & \multicolumn{4}{c|}{$\mathbf{Z}$} & \multicolumn{4}{c|}{$0$}\tabularnewline
\hline
2 & \multicolumn{4}{c|}{$0$} & \multicolumn{4}{c|}{$\mathbf{Z}$}\tabularnewline
\hline
$\vdots$ & \multicolumn{4}{c|}{$\vdots$} & \multicolumn{4}{c|}{$\vdots$}\tabularnewline
\hline
\hline
 & \multicolumn{8}{c|}{Real case}\tabularnewline
\hline
 & AI & D & AII & C & BDI & DIII & CII & CI\tabularnewline
\hline
$p\diagdown i$ & $\mathit{1}$ & $\mathit{3}$ & $\mathit{5}$ & $\mathit{7}$ & $\mathit{2}$ & $\mathit{4}$ & $\mathit{6}$ & $\mathit{8}$\tabularnewline
\hline
1 & $0$ & $\mathbf{Z}$ & $\mathbf{Z}^{(2)}_{2}$ & $2\mathbf{Z}$ & $0$ & $\mathbf{Z}^{(1)}_{2}$ & $0$ & $0$\tabularnewline
\hline
2 & $0$ & $0$ & $\mathbf{Z}^{(1)}_{2}$ & $0$ & $0$ & $\mathbf{Z}$ & $\mathbf{Z}^{(2)}_{2}$ & $2\mathbf{Z}$\tabularnewline
\hline
3 & $2\mathbf{Z}$ & $0$ & $\mathbf{Z}$ & $\mathbf{Z}^{(2)}_{2}$ & $0$ & $0$ & $\mathbf{Z}^{(1)}_{2}$ & $0$\tabularnewline
\hline
4 & $0$ & $0$ & $0$ & $\mathbf{Z}^{(1)}_{2}$ & $2\mathbf{Z}$ & $0$ & $\mathbf{Z}$ & $\mathbf{Z}^{(2)}_{2}$\tabularnewline
\hline
5 & $\mathbf{Z}^{(2)}_{2}$ & $2\mathbf{Z}$ & $0$ & $\mathbf{Z}$ & $0$ & $0$ & $0$ & $\mathbf{Z}^{(1)}_{2}$\tabularnewline
\hline
6 & $\mathbf{Z}^{(1)}_2$ & $0$ & $0$ & $0$ & $\mathbf{Z}^{(2)}_{2}$ & $2\mathbf{Z}$ & $0$ & $\mathbf{Z}$\tabularnewline
\hline
7 & $\mathbf{Z}$ & $\mathbf{Z}^{(2)}_{2}$ & $2\mathbf{Z}$ & $0$ & $\mathbf{Z}^{(1)}_{2}$ & $0$ & $0$ & $0$\tabularnewline
\hline
8 & $0$ & $\mathbf{Z}^{(1)}_{2}$ & $0$ & $0$ & $\mathbf{Z}$ & $\mathbf{Z}^{(2)}_{2}$ & $2\mathbf{Z}$ & $0$\tabularnewline
\hline
$\vdots$ & $\vdots$ & $\vdots$ & $\vdots$ & $\vdots$ & $\vdots$ & $\vdots$ & $\vdots$ & $\vdots$\tabularnewline
\hline
\end{tabular}
\caption{Classification of Fermi surfaces. $p$ is the codimension of an FS and $i$ is the index of symmetry classes. 
\label{tab:periodicity table}}
\end{table}


Let us look at the most general case, $i.e.$, the class A, in the absence of any
symmetry. The basic idea to classify FSs in terms of the topology
may be learnt from this case, which was already discussed in Ref.\cite{Horava}. We here follow it for a pedagogical
purpose. The Green's function can be written as
\begin{equation}
G(\omega,\mathbf{k})=\frac{1}{i\omega-\mathcal{H}(\mathbf{k})},\label{eq:Green's function}
\end{equation}
which is regarded as an N-dimensional matrix. FSs are defined
to be connected to the zero energy. Formulating topological charges in terms of the Green's function has an advantage when handling interacting systems, which is addressed in \cite{Order-parameter of TI,Horava,Interacting}. 
Generally the FS consists of branches
of compact manifolds in $\mathbf{k}$-space. For a specific branch
with dimension $d-p$,  a $p$-dimensional sphere
$S^{p}$ can always be picked up from $(\omega,\mathbf{k})$-space to enclose this branch in its transverse dimension. $G^{-1}(\omega,\mathbf{k})$ is nonsingular for each $(\omega,\mathbf{k})$
restricted on the $S^{p}$; in other words, it is a member of $GL(N,\mathbf{C})$.
Then $G^{-1}(\omega,\mathbf{k})$ restricted on the $S^{p}$ can be
regarded as a mapping from $S^{p}$ to $GL(N,\mathbf{C})$. As all
these mappings can be classified by the homotopy group $\pi_{p}(GL(N,\mathbf{C}))$,
$G^{-1}(\omega,\mathbf{k})$ on the $S^{p}$ is in a certain homotopy
class. In the so-called stable regime, $N>\frac{p}{2}$, $\pi_{p}(GL(N,\mathbf{C}))$
satisfies the Bott periodicity\cite{Nakahara},
\[
\pi_{p}\left(GL(N,\mathbf{C})\right)\cong\begin{cases}
\mathbf{Z} & p\in odd\\
0 & p\in even
\end{cases}.
\]
Since physical systems are always in the stable regime, only the mappings
on $S^{p}$ for FSs with the odd codimension $p=2n+1$ can
be in a nontrivial homotopy class, while ones for the even codimension
$p$ are always trivial. In other words, FSs with the even
codimension $p$ are always trivial and vulnerable against weak perturbations,
while ones with odd codimension can be classified by $\pi_{p}(GL(N,\mathbf{C}))\cong\mathbf{Z}$
and its stability is topologically protected
against any weak perturbations for the nontrivial one. We emphasize that since the class
A is not subject to any symmetry, the perturbation can be quite
arbitrary. In addition, the concrete shape of the $S^{p}$ is irrelevant
in principle, and the only requirement is that it is an orientable
compact submanifold in $(\omega,\mathbf{k})$-space with dimension
$p$, which is distinctly different from the cases of reality symmetries. The
homotopy number of a mapping for an FS is named as the
topological charge of the FS, which can be calculated from
the following formula,
\begin{equation}
N_{p}=C_{p}\int_{S^{p}}\mathbf{tr}\left(G\mathbf{d}G^{-1}\right)^{p}\label{eq:homotopic number}
\end{equation}
with
\[
C_{p}=-\frac{n!}{(2n+1)!(2\pi i)^{n+1}}.
\]

The $\mathbf{Z}$-type of FSs in the class A is
thus obtained. Once an FS has nontrivially this topological
charge, it can survive under any weak perturbations
in the absence of any symmetry\cite{Hubbard}. But the topologically trivial FSs
can be gapped due to the perturbations. Both cases can be seen in
$^{3}He$\cite{Volovik-Book}, where two Fermi points in $^{3}He$-$A$ phase are topologically protected with charge $\pm2$ individually, while the Fermi line in the
planar phase cannot be topologically charged because its codimension is even. The FS of another simple model Hamiltonian $\mathcal{H}=\mathbf{g}(\mathbf{k})\cdot\mathbf{\sigma}$ belongs also to this topological class\cite{Supp}, where $\sigma_i$ are Pauli matrices.

We now turn to consider the case when the system has a chiral
symmetry denoted by $\mathcal{K}$, $i.e.$,
\[
\left\{ \mathcal{H},\mathcal{K}\right\} =0.
\]
Systems with this symmetry have a crucial property: for any eigenstate
$|\alpha\rangle$ of $H$ with the energy $E_{\alpha}$, $K|\alpha\rangle$
is also an eigenstate of $H$ but with the energy $-E_{\alpha}$. The
Hamiltonian associated with this symmetry can alway be diagonalized in $\mathbf{k}$-space
on the basis that diagonalizes $\mathcal{K}$ and thus can be written as
\[
h(\mathbf{k})=\begin{pmatrix} & u(\mathbf{k})\\
u^{\dagger}(\mathbf{k})
\end{pmatrix}.
\]
On the $S^{p}$ that is chosen to enclose an FS of dimension $d-p$,
$h(\mathbf{k})$ also takes this form, which makes the topological charge
defined in Eq(\ref{eq:homotopic number}) always vanishing. However
$u(\mathbf{k})$ can be regarded as a mapping from $S^{p-1}$(set
$\omega=0$) to $GL(N/2,\mathbf{C})$, which may be in a nontrivial
homotopy class of $\pi_{p-1}(GL(N/2),\mathbf{C})$. In this sense, there exist
nontrivial FSs for the even number of codimension, $i.e.$, $p=2n$. The
homotopy number may be calculated as
\[
\nu_{p}=C_{p-1}\int_{S^{p-1}}\mathbf{tr}\left(u(\mathbf{k})\mathbf{d}u^{-1}(\mathbf{k})\right)^{p-1}.
\]
As the homotopic number is real, it can also be calculated by $u^{\dagger}(\mathbf{k})$.
Thus the homotopic number can also be expressed by Green's function:
\begin{equation}
\nu_{p}=\frac{C_{p-1}}{2}\int_{S^{p-1}}\mathbf{tr}\left(\mathcal{K}\left(G\mathbf{d}G^{-1}\right)^{p-1}|_{\omega=0}\right).\label{eq:general chiral charge}
\end{equation}
The homotopic number is referred to as the chiral topological charge of
the FS.

For this $\mathbf{Z}$-type topological charge in the class AIII, if
it is nontrivial, the FS is stable
against any perturbations that do not break  the CS. However the
topological protection is not as stable as that induced by Eq.(\ref{eq:homotopic number}),
because  the FS is gapped if ever the CS is broken. For instance,   the honeycomb lattice model  has two Dirac cones with the low-energy effective Hamiltonian $\mathcal{H}_\alpha=k^{\pm}_x\sigma_1\pm k_y^{\pm}\sigma_2$, respectively, with  $\sigma_3$ representing the chiral symmetry that is the sublattice symmetry here \cite{Haldane model}. The two Dirac cones have $\nu_2=\pm1$ respectively\cite{Supp}, and thus they are topologically stable as far as  this chiral symmetry is preserved.  More remarkably, according to our present theory, it is noted that the stability of FSs in this model was recently  verified in an ultra-cold atom experiment\cite{Dirac Points}.  In addition, it was also seen clearly from this experiment that if the sublattice symmetry is broken, a gap may be opened in the whole Brillouin zone. It is noted that this type of topological charge was also seen in a superconducting system\cite{SConductor}. Another interesting model Hamiltonian $\mathcal{H}=(k^2_x-k^2_y)\sigma_1\pm2k_xk_y\sigma_2$ also gives out this type of topological charge $\nu_2=\pm2$ \cite{Supp}.


 
When TRS and PHS are considered, many $\mathbf{Z}$-type topological charges vanish. As for the first descendant ($i.e.\,\,\mathbf{Z}^{(1)}_2$) of a $\mathbf{Z}$-type  in a non-chiral real case ,  $e.g.$
 the case of codimension 2 that is labeled one row above that of codimension 3 ($\mathbf{Z}$-type) in the class AII in Tab.{[}\ref{tab:periodicity table}{]},
the Green's function restricted on the chosen $S^{p}$ can be classified by a $\mathbf{Z}_2$-type topological charge.
The key idea is that $G(\omega,\mathbf{k})|_{S^p}$ can be continuously extended
to $\tilde{G}(\omega,\mathbf{k};u)$ on the $(p+1)$-dimensional disk  by introducing
an auxiliary parameter $u$ (ranging from $0$ to $1$) with the two
requirements\label{sub: requirement I}: $i)$ $\tilde{G}(\omega,\mathbf{k};0)|_{S^{p}}=G(\omega,\mathbf{k})|_{S^{p}}$,
and $ii)$ $\tilde{G}(\omega,\mathbf{k};1)|_{S^{p}}$ is a diagonal
matrix, such that $\tilde{G}_{\alpha\alpha}=\left(i\omega-\Delta\right)^{-1}$for
empty bands and $\tilde{G}_{\beta\beta}=(i\omega+\Delta)^{-1}$ for
occupied bands, where $\Delta$ is a positive constant\cite{Order-parameter of TI,Xiao-Liang PRB}.
The validity of the extension is based on the fact that $G(\omega,\mathbf{k})$
restricted on $S^{p}$ is always trivial in the homotopic sense in
this case. This topological charge in terms of Green's function is
formulated as
\begin{equation}
N_{p}^{(1)}=C'_{p}\int_{S^{p}}\int_{0}^{1}du\;\mathbf{tr}\left(\left(G\mathbf{d}G^{-1}\right)^{p}G\partial_{u}G^{-1}\right)\mod\:2\label{eq:Z_2 son for non-chiral case}
\end{equation}
with
\[
C_{p}^{'}=-\frac{2(p/2)!}{p!(2\pi i)^{(p/2)+1}},
\]
where ``$\sim$'' has been dropped for brevity. The topological
charge is defined in a similar way to the WZW-term in quantum field theory\cite{WZW-term}
and the $\mathbf{Z}_{2}$ character comes from the fact that two different
extensions differ to an even integer, as demonstrated in Ref.\cite{Order-parameter of TI}. As a concrete illustration, we here exemplify this $\mathbf{Z}_2$-type topological charge by a model Hamiltonian defined in a two spatial dimension: $\mathcal{H}(\mathbf{k})=k_x\sigma_1+k_y\sigma_2+(k_x+k_y)\sigma_3$, which  has only a TRS with $C=i\sigma_2$ and $\eta=-1$ according to Eq.(\ref{eq:reality transformation}),  and thus belongs to the class AII in Tab.{[}\ref{tab:classification}{]} and corresponds to the case of $R(2,5)$ in Tab.{[}\ref{tab:periodicity table}{]}. The corresponding FS at the origin in the $(\omega,\mathbf{k})$-space 
is found to have a nontrivial topological charge $N_{2}^{(1)}=1 $ from Eq.(\ref{eq:Z_2 son for non-chiral case}) \cite{Supp} . 

For the second descendant with the codimension $p$ ($i.e.\,\,\mathbf{Z}^{(2)}_2$), $e.g.$ the case
of codimension 3 with two rows above that of codimension 5 in the class C in Tab.{[}\ref{tab:periodicity table}{]},
 a $\mathbf{Z}_{2}$-type topological charge can also be defined on the
chosen $S^{p}$. To define the $\mathbf{Z}_{2}$-type topological charge,
the $G(\omega,\mathbf{k})|_{S^{p}}$ is smoothly extended to a two-dimensional
torus $T^{2}$ parameterized by the two auxiliary parameters $u$
and $v$ (both ranging from $-1$ to $1$ ) with the three requirements:
$i)$ $\tilde{G}(\omega,\mathbf{k};0,0)|_{S^{p}}=G(\omega,\mathbf{k})|_{S^{p}}$; and
$ii)$ $\tilde{G}(\omega,\mathbf{k};u,v)|_{S^{p}}=\epsilon_{c}C\tilde{G}^{T}(\omega,-\mathbf{k};-u,-v)|_{S^{p}}C^{-1}$,
referring to Eq.(\ref{eq:reality transformation}); $iii)$$\tilde{G}(\omega,\mathbf{k};1,1)|_{S^{p}}$
corresponds to a trivial system, such as $\tilde{G}_{\alpha\alpha}=\left(i\omega-\Delta\right)^{-1}$for
empty bands and $\tilde{G}_{\beta\beta}=(i\omega+\Delta)^{-1}$ for
occupied bands. The corresponding topological
charge is defined as\cite{Xiao-Liang PRB,Order-parameter of TI}
\begin{equation}
N_{p}^{(2)}=C{}_{p+2}\int_{S^{p}\times T^{2}}\:\mathbf{tr}\left(G\mathbf{d}G^{-1}\right)^{p+2}\mod2,\label{eq:Z_2 grandson fro non-chiral case}
\end{equation}
where ``$\sim$'' has been dropped and
the $C_{p+2}$ is defined in Eq.(\ref{eq:homotopic number}). As a topological charge of  FSs, its physical meaning is analogous to that
of Eq.(\ref{eq:Z_2 son for non-chiral case}).


Similar to the non-chiral case, a $\mathbf{Z}$ chiral topological
charge is associated with two descendants: son and grandson. The $\mathbf{Z_{2}}$
topological charge of the son is originated from the chiral topological
charge defined in Eq.(\ref{eq:general chiral charge}) in the same
spirit of that Eq.(\ref{eq:Z_2 son for non-chiral case}) is originated from
Eq.(\ref{eq:homotopic number}), which is given by
\begin{equation}
\begin{split}
\nu_{p}^{(1)}=&\frac{C'_{p-1}}{2}\int_{S^{p-1}}\int_{0}^{1}du\;\\
&\mathbf{tr}\left(\mathcal{K}\left(G\mathbf{d}G^{-1}\right)^{p-1}G\partial_{u}G^{-1}|_{\omega=0}\right)\mod\:2,\label{eq:Chiral son charge}
\end{split}
\end{equation}
where $\mathcal{K}$ is the matrix to represent the chiral symmetry,
$S^{p-1}$ is the chosen $S^{p}$ restricted on $\omega=0$, and $G$
is extended by an auxiliary parameter ranging from $0$ to $1$ with
the same requirements for introducing Eq.(\ref{eq:Z_2 son for non-chiral case}). The $\mathbf{Z}_{2}$ topological charge for the grandson can also
be defined in the same spirit of writing out Eq.(\ref{eq:Z_2 grandson fro non-chiral case}).
We extend the $G(\omega,\mathbf{k})|_{S^{p}}$ to a two-dimensional
torus $T^{2}$ parameterized by two auxiliary parameters $u$ and
$v$ (both ranging from $-1$ to $1$) with the four requirements:
the first three requirements are the same as those of the non-chiral
counterpart in Eq.(\ref{eq:Z_2 grandson fro non-chiral case}); while the
fourth one is that the chiral symmetry is preserved on $T^{2}$, which
also permits that either of reality symmetries is applicable in the second
requirement. The $\mathbf{Z}_{2}$ topological charge is written as
\begin{equation}
\nu_{p}^{(2)}=\frac{C_{p+1}}{2}\int_{S^{p-1}\times T^{2}}\:\mathbf{tr}\left(\mathcal{K}\left(G\mathbf{d}G^{-1}\right)^{p+1}|_{\omega=0}\right)\mod2,\label{eq:chiral grandson charge}
\end{equation}
where $C_{p+1}$ is defined in Eq.(\ref{eq:homotopic number}).

The physical meanings of topological charges for the eight classes with TRS and/or PHS are elaborated as
follows. The FS(s) with a certain codimension is always distributed centrosymmetrically, since either TRS or PHS relates  $\mathbf{k}$ to $-\mathbf{k}$. Thus there exist two possibilities:  $i)$ the FS with codimension $p$ resides at the origin in $(\omega,\mathbf{k})$-space; $ii)$the FS(s) is centrosymmetric outside of the origin.  For the first case, we can choose an $S^p$ in $(\omega,\mathbf{k})$-space to enclose the FS, and use the corresponding formula to calculate the topological charge. If the topological charge is nontrivial, the FS is stable against perturbations provided the corresponding symmetries are preserved.
For the second case, the FS(s) is usually spherically distributed, so two $S^{p}$s can be chosen
to sandwich the FS(s) in its transverse dimension. The difference of the topological charges calculated from the two $S^p$s is the topological charge of the FS(s). If it is nontrivial, the FS(s) is topologically stable when the corresponding symmetries are preserved.

Before concluding this paper, we wish to emphasize that the topological charges of FSs addressed here are closely connected to topological insulators/superconductors\cite{Zhao1}, and therefore this work may provide a new and deep insight for studying them.


To conclude, FSs 
in all of the ten symmetry
classes have been classified appropriately in terms of topological charges.
It has been revealed that when an FS is associated with a nontrivial topological charge,  this FS is topologically protected
by the corresponding symmetry.

\begin{acknowledgments}
We thank G. E. Volovik for helpful discussions. This work was supported
by the GRF (HKU7058/11P),  the CRF (HKU8/11G) of Hong Kong, the URC fund of HKU, and 
the SKPBR of China (Grant No. 2011CB922104).
\end{acknowledgments}

\section{Supplemental Material}
As representative examples, we illustrate below how to calculate the topological charges of  Fermi surfaces  in the models with different classes of topological classification.

\section{A. Calculation of the $Z$-type topological charge in the class A}
We first look at a simple three-dimensional model Hamiltonian given by
\begin{equation*}
\mathcal{H}(\mathbf{k})=\mathbf{g}(\mathbf{k})\cdot\mathbf{\sigma}\,\,\, or\,\,\,G^{-1}=i\omega-\mathbf{g}(\mathbf{k})\cdot\mathbf{\sigma},
\end{equation*}
where $\sigma_i$ (i=1,2,3) are Pauli matrices and $\mathbf{g}(\mathbf{k})$ is equal to zero only if $\mathbf{k}=0$, $i.e.$, the Fermi surface is a point at the origin in $(\omega,\mathbf{k})$-space. Here, it is also assumed that this model with a given $\mathbf{g}(\mathbf{k})$ has no discrete symmetries, so that it belonged to the class A. For convenience, an $S^2$ is chosen in $\mathbf{k}$-space, and then is translated along the $\omega$-direction to form an infinitely long cylinder enclosing the Fermi surface. Choosing $(\theta,\phi)$ as the coordinates of the $S^2$, we parametrize the three-dimensional cylinder as $x^\mu=(\omega,\theta,\phi)$. According to Eq.(4), the corresponding topological charge is given by
\begin{equation*}
N_3=\frac{1}{24\pi^2}\mathbf{tr}\int d^3 x\,\,\epsilon^{\mu\nu\rho}\,G\partial_\mu G^{-1}G\partial_\nu G^{-1}G\partial_\rho G^{-1}.
\end{equation*}
Substituting $G^{-1}$ into the equation and using $\mathbf{tr}(\sigma_i\sigma_j\sigma_k)=2i\epsilon_{ijk}$, we have 
\begin{equation*}
N_3=\frac{1}{4\pi^2}\epsilon^{ij}\epsilon_{nml}\int d\omega d\theta d\phi \frac{1}{(\omega^2+|\mathbf{g}|^2)^2} g^n\partial_i g^m \partial_j g^l.
\end{equation*}
Using 
\begin{equation*}
\int_{-\infty}^{\infty} d\omega\,\,\frac{1}{(\omega^2+|\mathbf{g}|^2)^2}=\frac{\pi}{2|\mathbf{g}|^3},
\end{equation*}
we can find
\begin{equation*}
N_3=\frac{1}{4\pi}\int d\theta d\phi\frac{1}{|\mathbf{g}|^3}\mathbf{g}\cdot(\partial_\theta \mathbf{g}\times\partial_\phi\mathbf{g}). 
\end{equation*}
Defining a unit vector $\hat{\mathbf{g}}=\mathbf{g}/|\mathbf{g}|$, the above expression can be further simplified as
\begin{equation*}
N_3=\frac{1}{4\pi}\int d\phi d\theta\,\,\hat{\mathbf{g}}\cdot(\partial_\theta \hat{\mathbf{g}}\times\partial_\phi\hat{\mathbf{g}}),
\end{equation*}
which is just the winding number from $S^2$ to $S^2$. A nonzero $N_3$ implies the topological stability of the Fermi surface. As a very simple example, $\mathcal{H}=\pm\mathbf{k}\cdot \sigma$, so that $\hat{\mathbf{g}}=\pm(\sin\theta\cos\phi,\sin\theta\sin\phi,\cos\theta)$. The corresponding $Z$-type topological charge is
\[
N_3=\pm\frac{1}{4\pi}\int_0^{2\pi}d\phi \int_0^\pi d\theta \sin\theta=\pm1.
\]
In other words, Weyl fermions have topological charge equal to their chiralities [1].
\section{B. Evaluation of the $\mathbf{Z}$-type topological charge for a physical model in the class AIII}

Let us consider such an effective two-dimensional model Hamiltonian given by
\[
\mathcal{H_{\pm}}(\mathbf{k})=k^{\pm}_{x}\sigma_{1}\pm k^{\pm}_{y}\sigma_{2},
\]
where $\pm$ denote the two Dirac cones. 
This model has a chiral symmetry, $i.e.,$ $\{\mathcal{H},\sigma_{3}\}=0$,
and thus it belongs to the class AIII. The Fermi surface is obviously
the origin point in the $(\omega,\mathbf{k})$-space. We here choose an
$S^{1}$ enclosing the Fermi surface, and calculate the topological
charge according to Eq.(5): 
\[
\nu_{2}=\frac{1}{4\pi i}\int_{0}^{2\pi}d\phi\:\mathbf{tr}\left(\sigma_{3}\mathcal{H}^{-1}(\phi)\partial_{\phi}\mathcal{H}(\phi)\right),
\]
where $\mathcal{H}(\phi)=k\cos\phi\sigma_{1}\pm k\sin\phi\sigma_{2}.$ After simplifying the above expression, we obtain
\[
\nu_{2}=\frac{1}{4\pi i}\int_0^{2\pi} d\phi\,\,\,(\pm2i)=\pm1.
\]
Thus the Fermi surface is topologically stable as far as  the chiral symmetry is preserved. 

Another interesting but a little bit more complicated model in this class reads 
\[
\mathcal{H}=(k^2_x-k^2_y)\sigma_1\pm2k_xk_y\sigma_2,
\]
which may be simulated by cold atom experiments. We choose a square path with the side length $2a$ centered at the origin. Since every side is equivalent to each other, we focus on the right side, where
\[
\mathcal{H}=\left(\begin{array}{cc}
0 & (a\mp ik_y)^2\\
(a\pm ik_y)^2 & 0
\end{array}\right).
\]
Thus
\[
\nu_{2} = \frac{1}{4\pi i}\mathbf{tr}\oint_{C}dl\:\sigma_{3}\mathcal{H}^{-1}\partial_{l}\mathcal{H}(\mathbf{k})=\pm \frac{2a}{\pi}\int_{-a}^{a}\frac{dk_{y}}{a^{2}+k_{y}^{2}}=\pm2.
\]
The doubly charged  Fermi surface in this model is also topologically stable.

\section{C. An example of $\mathbf{Z}_2$-type topological charges in the Class AII}

We here address a two-dimensional model Hamiltonian as
\[
\mathcal{H}(\mathbf{k})=k_{x}\sigma_{1}+k_{y}\sigma_{2}+(k_{x}+k_{y})\sigma_{3}
\]
This Hamiltonian has a time-reversal symmetry(TRS) with sign $-$,
$i.e.$,
\[
\mathcal{H}(\mathbf{k})=\sigma_{2}\mathcal{H}^{T}(-\mathbf{k})\sigma_{2}^{-1},\quad\sigma_{2}^{T}=-\sigma_{2}.
\]
It is straightforward to check that this Hamiltonian has neither the chiral symmetry
nor the particle-hole symmetry, and so it belongs to the class AII. The Fermi
surface of this model is the point $\mathbf{k}=0$ with codimension
$2$. According to Table{[}II{]}, we can calculate its $\mathbf{Z_{2}}$
topological charge from Eq.(6). In order to do this, we choose an $S^{2}$ in $(\omega,\mathbf{k})$-space enclosing
the origin, and make the following continuous extension of the Green's
function restricted on the $S^{2}$:
\begin{eqnarray*}
G^{-1} & = & i\omega-\Delta\sigma_{3}\sin\theta\\
 &  & -\left[k\cos\phi\sigma_{1}+k\sin\phi\sigma_{2}+k(\cos\phi+\sin\phi)\sigma_{3}\right]\cos\theta ,
\end{eqnarray*}
where  $\phi\in[0,2\pi]$ parametrizes the circle in $\mathbf{k}$-space enclosing the Fermi surface,  $\theta\in[0,\pi/2]$ is the parameter for extension, and $\Delta$ is a positive
constant. Substituting this Green's function into Eq.(6) and using a result in Sec. A, we obtain
\[
N_{2}^{(1)}=\frac{1}{2\pi}\int_{0}^{2\pi}d\phi\int_{0}^{\frac{\pi}{2}}d\theta\;\hat{\mathbf{g}}\cdot\left(\partial_{\theta}\hat{\mathbf{g}}\times\partial_{\phi}\hat{\mathbf{g}}\right),
\]
with 
$$\mathbf{g}=\left(k\cos\phi\cos\theta\,,k\sin\phi\cos\theta,k(\cos\phi+\sin\phi)+\Delta\sin\theta\right).$$
For simplicity, setting $\alpha=k=\Delta=1$, we obtain
\[
N_2^{(1)}=\frac{1}{2\pi}\int_{0}^{2\pi}d\phi\int_{0}^{\frac{\pi}{2}}d\theta\,\,\, \frac{n(\theta,\phi)}{m(\theta,\phi)} =1,
\]
where
\[
n(\theta,\phi)=\cos\theta
\]
and
\begin{eqnarray*}
m(\theta,\phi)&=&[\sin^2\theta+\sin(2\theta)(\cos\phi+\sin\phi)\\
                      & &+\cos^2\theta(2+\sin(2\phi))]^{\frac{3}{2}}
                      \end{eqnarray*}
It is found that
$N_{2}^{(1)}=1 $ in this case, and thus the Fermi surface is protected
by a nontrivial $\mathbf{Z_{2}}$ topological charge as far as this TRS is preserved. Finally, it is worth pointing out that the present model may be experimentally simulated with ultra-cold atoms, and thus our prediction of this $\mathbf{Z_{2}}$ topological feature can be tested by future experiments.


\begin{thebibliography}{References}

\bibitem{Volovik-Book}G. E. Volovik, \textit{The Universe in a Helium Droplet}
(Clarendon, Oxford, 2003).

\bibitem{Volovik Vacuum}G. E. Volovik, arXiv:1111.4627 {[}hep-ph{]} (2011).

\bibitem{Horava}P. Ho\v{r}ava, Phys. Rev. Lett \textbf{95}, 016405
(2005).

\bibitem{Ramdom Matrix 0}M. R. Zirnbauer, J. Math. Phys. \textbf{37},
4986 (1996).

\bibitem{Ramdom Matrix I}A. Altland and M. R. Zirnbauer, Phys. Rev.
B \textbf{55}, 1142 (1997).

\bibitem{TI Classification III}A. P. Schnyder, S. Ryu, A. Furusaki,
and A. W. W. Ludwig, Phys. Rev. B \textbf{78}, 195125 (2008).

\bibitem{Ramdom Matrix II}D. Bernard and A. LeClair, J. Phys. A:
Math. Gen. \textbf{35}, 2555 (2002).


\bibitem{TI Classification I}S. Ryu, A. P. Schnyder, A. Furusaki
and A. W. W Ludwig, New J. Phys. \textbf{12}, 065010 (2010).

\bibitem{Tl Classification II}A. P. Schnyder, S. Ryu, A. Furusaki,
and A. W. W. Ludwig, AIP Conf. Proc. \textbf{1134}, 10 (2009).

\bibitem{note1}Noting that the Fermi surface may consist of several disconnected parts, each having different topological characteristics, for simplicity and without loss of generality, we here focus only on one branch.

\bibitem{note}$\mathbf{Z}$ and 2$\mathbf{Z}$ are regarded to be the same type of topological charge given by Eq.[\ref{eq:homotopic number}] (Eq.[\ref{eq:general chiral charge}]) in the chiral (non-chiral) real case, while $\mathbf{Z}^{(1)}_2$ and $\mathbf{Z}^{(2)}_2$ are different types.


\bibitem{Nakahara}M. Nakahara, \textit{Geometry, Topology and Physics} (Adam Hilger, Bristol, 1990).

\bibitem{Order-parameter of TI} Z. Wang, X. L. Qi and S. C. Zhang,
Phys. Rev. Lett. \textbf{105}, 256803 (2010).

\bibitem{Interacting}Z. Wang and S. C. Zhang, Phys. Rev. X \textbf{2}, 031008 (2012).

\bibitem{Hubbard}Noting that for the square lattice Hubbard model at half-filling with a very large U (no continuum approximation could be used), a seemingly perturbative effect of adding one electron or hole will significantly shift the ground state of the system, such that it is effectively non-perturbative and thus may not be taken care of from the present topological analysis .

\bibitem{Supp} Supplementary Material.


\bibitem{Haldane model}F. D. M. Haldane, Phys. Rev. Lett. \textbf{61},
2015 (1988).

\bibitem{Dirac Points}L. Tarruell, D. Greif, T. Uehlinger, G. Jotzu
and T. Esslinger, Nature \textbf{483}, 302 (2012).

\bibitem{SConductor}B. Beri, Phys. Rev. B \textbf{81}, 134515 (2010).

\bibitem{Xiao-Liang PRB}X. L. Qi, T. L. Hughes, and S. C. Zhang,
Phys. Rev. B \textbf{78}, 195424 (2008).

\bibitem{WZW-term}E. Witten, Nucl. Phys. B. \textbf{223}, 422 (1983).


\bibitem{Zhao1} Y. X. Zhao and Z. D. Wang, arXiv:1305.1251 (2013); arXiv:1305.3791 (2013).


\end{thebibliography}
\end{document}